\documentclass[journal=nanoletters,manuscript=article]{achemso}
\usepackage[T1]{fontenc}
\usepackage[utf8]{inputenc}
\usepackage{graphicx}
\usepackage{pstricks}
\usepackage{array}
\usepackage{natbib}
\usepackage{amsmath}
\usepackage{pifont}
\usepackage{dsfont}
\usepackage{multirow}
\usepackage{amssymb}
\usepackage{wasysym}
\usepackage[titletoc,toc,title]{appendix}
\usepackage{booktabs}
\usepackage{colortbl}
\usepackage{xcolor}
\usepackage{mathtools}
\usepackage{gensymb}
\usepackage{breqn}

\usepackage{fancyhdr}

\def\be{\begin{equation}}
\def\ee{\end{equation}}

\newcommand{\eq}[1]{Equation~(\ref{#1})}

\def\GB{\mathrm{_{GB}}}

\author{Benoit Gaury}
\affiliation[National Institute for Standards and Technology]
{Center for Nanoscale Science and Technology, National Institute for Standards and Technology, Gaithersburg, MD 20899, USA}
\alsoaffiliation[University of Maryland]
{Maryland NanoCenter, University of Maryland, College
Park, MD 20742, USA}
\author{Paul M. Haney}
\email{paul.haney@nist.gov}
\affiliation[National Institute for Standards and Technology]
{Center for Nanoscale Science and Technology, National Institute for Standards and Technology, Gaithersburg, MD 20899, USA}

\title[An \textsf{achemso} demo]
  {Quantitative theory of the grain boundary impact on the open-circuit voltage of polycrystalline solar cells}

\providecommand{\mykeywords}[1]{\textbf{\textit{Keywords---}} #1}

\begin{document}
\mykeywords{photovoltaics, grain boundary, thin film, polycrystalline semiconductor, carrier defect recombination}

\begin{abstract}
Thin film polycrystalline photovoltaics are a mature, commercially-relevant technology.  However, basic questions persist about the role of grain boundaries in the performance of these materials, and the extent to which these defects may limit further progress.  In this work, we first extend previous analysis of columnar grain boundaries to develop a model of the recombination current of ``tilted'' grain boundaries.  We then consider systems with multiple, intersecting grain boundaries and numerically determine the parameter space for which our analytical model accurately describes the recombination current.  We find that for material parameters relevant for thin film photovoltaics, our model can be applied to compute the open-circuit voltage of materials with networks of inhomogeneous grain boundaries.
This model bridges the gap between the distribution of grain boundary properties observed with nanoscale characterization and their influence on the macroscale device open-circuit voltage.
\end{abstract}


\maketitle

\bigskip
\noindent

Polycrystalline materials possess an abundance of extended crystallographic
defects in the form of grain boundaries, which are typically harmful to device
performance. However, recent development in thin-film  photovoltaics have
led to surprisingly high efficiencies given the large densities of grain
boundaries.~\cite{kumar2014physics} The efficiency records were
obtained mostly by improvements in light absorption and collection of
photogenerated carriers.~\cite{geisthardt2015status}
Increasing the open-circuit voltage (currently at $\approx 880~{\rm mV}$ in
polycrystalline CdTe \cite{geisthardt2015status, gloeckler2013cdte}) has proven to be more difficult.
Two groups recently reported~\cite{Zhao2016, Burst2016} single crystal CdTe solar cells with open-circuit voltages above $1~\rm
V$, suggesting that grain boundaries may be an important source of
recombination and reduce the open-circuit voltage of polycrystalline solar
cells \cite{sites1998losses}. While grain boundaries are a predominant source of defects in thin film
photovoltaics, a precise understanding of grain boundary recombination and its
impact on performance remains uncertain and controversial.~\cite{major2016grain}  A primary difficulty in experimentally determining the effect of grain boundaries is that modifying grain structure typically changes other important material properties~\cite{major2016grain} (as in the studies of Ref.~\citenum{Zhao2016, Burst2016}).  Theoretical models are suited to provide guidance in this case, as models afford the freedom to independently vary material and system parameters.  The simplest models
of grain boundaries~\cite{Card1977, Fossum1980,green1996bounds,
edmiston1996improved} used for analyzing polycrystalline Si are inconsistent
with the high efficiency of thin film photovoltaics, indicating the need for
more sophisticated approaches.

On the experimental side, there has been recent substantial progress in characterizing grain boundaries.
Nanoscale imaging and spectroscopy can, in some circumstances, reveal the full three-dimensional, chemically resolved atomic structure of grain boundaries.~\cite{Wang2011,Sun2016}  Knowledge of atomic
structure enables first principles calculations of the electronic structure
of certain ideal grain boundaries, identifying defect energy levels and charge
states.~\cite{Yanfa2015} Direct measurements of electrical properties of
individual grain boundaries using high resolution techniques yield qualitative
insights (such as the sign of grain boundary defect charge~\cite{yoon2013local,Jiang2004}), although quantitative interpretation
of these measurements remains a challenge. Nevertheless, even perfect knowledge
of grain boundary electrical properties would not suffice to determine their
impact on important figures of merit, such as the open-circuit voltage. This is
due to a gap on the theory side: so far no analytical relation connects grain
boundary properties of a realistic sample to its $V_{\rm oc}$. Here we provide this
previously missing component of the theory and demonstrate its validity for
material parameters typical of thin film solar cells. While the short circuit
current and the fill factor are also key elements of a solar cell efficiency, we
focus on the open-circuit voltage as it is the metric for which the largest
margin of improvement is available.~\cite{geisthardt2015status}

\begin{figure}[t]
\centering
 \includegraphics[width=0.49\textwidth]{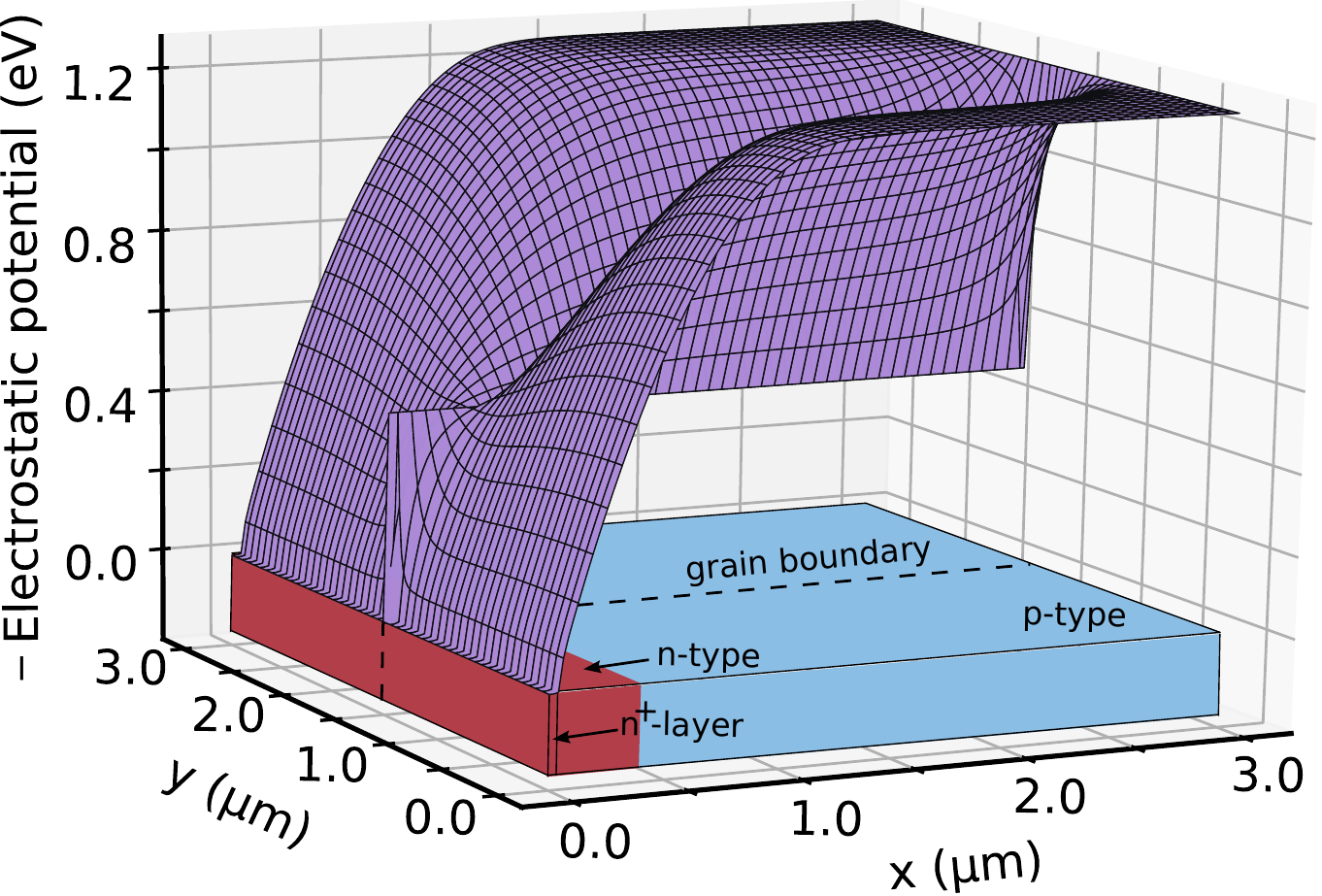}
 \caption{\label{pot} Thermal equilibrium electrostatic potential of a $pn^+$ junction
 containing a single charged grain boundary at $y=1.5~\mu\rm m$ (dashed line).}
\end{figure}

In a series of recent works,~\cite{Gaury2016GB, Gaury2017} we studied the
charge transport associated with isolated, columnar grain boundaries in thin
film solar cells consisting of $n^+p$ junctions ($p$-type absorber).
We obtained an approximate analytic solution for the grain boundary recombination current
under the conditions that the grain boundary is positively charged with large
defect density (so that the Fermi level is pinned to the defect neutrality level), and that the majority carrier
transport is sufficiently facile so that the quasi-hole Fermi level varies by less than the thermal voltage $V_T$ ($\approx 25~{\rm meV}$ at room temperature).
Under these circumstances, electrostatic screening leads to downward
band bending in the vicinity of the grain boundary (shown in Figure~\ref{pot}), which confines electrons near the grain boundary core.
We showed that in this case, the two-dimensional problem for the recombination can be mapped to an effective
one-dimensional problem for the motion of electrons along the
grain boundary. The dark recombination current of an isolated columnar grain boundary
versus voltage $V$ is shown to take the following general form~\cite{Gaury2017}
\be
    J\GB(V)=\lambda \frac{S}{2d} N e^{-E_a/k_BT} e^{qV/(nk_BT)},
    \label{form}
\ee
where $S$ is an effective surface recombination velocity, $\lambda$ is the characteristic length
over which recombination occurs, $d$ is the grain size, $N$ is an
effective density of states, $E_a$ is an activation energy, $n$ is the ideality
factor ($k_B$ is the Boltzmann constant and $T$ is the temperature).

The specific form of the parameters
depends on the type (i.e., majority carrier) of the grain boundary core.  There are 3 possible cases: 1. $n$-type, which
occurs when the band bending at the grain boundary is large enough to cause type inversion at the grain boundary core ({\it i.e.}
the Fermi level is closer to the conduction band at the grain boundary core), 2.  $p$-type, where we note that the assumption of downward band bending implies that the grain boundary core will always be less $p$-type than the bulk of the absorber.  For both $n$-type and $p$-type grain boundary cases, the majority carriers have a constant concentration along the entire length of the grain boundary.  The last case is: 3.  Neither $n$-type or $p$-type, a case to which we refer as ``high recombination''.  For this case, there are regions along the grain boundary core at which the electron and hole densities are similar in magnitude, and both carrier
densities vary along the length of the grain boundary. The specific expressions of the parameters entering Eq. (\ref{form}) are given in the Supporting
Information and in Ref.~\citenum{Gaury2016GB}.

\begin{figure}
\centering
\includegraphics[width=0.40\textwidth]{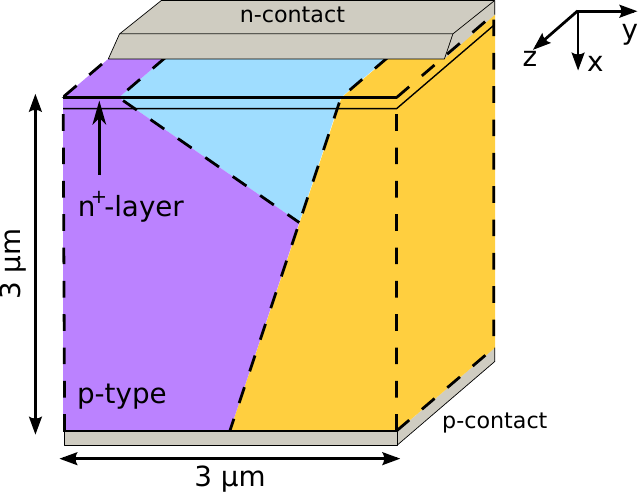} \caption{\label{geometry}
 Model system of a $pn^+$ junction containing several grains and grain boundaries (represented by the dashed lines).}
\end{figure}

In this work, we focus on microstructures with complex grain boundary topology, as depicted in
Figure~\ref{geometry}. We first extend our previous model to consider grain boundaries tilted
at an angle $\theta$ with respect to the $pn^+$ junction normal. Based on the physical picture of carrier recombination developed in previous works, we make a simple {\it ansatz} for the dependence of grain boundary recombination on $\theta$.
To demonstrate the validity of this {\it ansatz}, we make comparisons to 2-d numerical simulations performed with
the semiconductor modeling software Sesame \cite{gaury2018sesame}. We next analyze the carrier transport in {\it networks} of
non-columnar grain boundaries. We find that under similar assumptions leading to Eq. (\ref{form}),
the recombination of a particular grain boundary embedded within a network is
approximately equal to the recombination of the same grain boundary in
isolation.
The total dark recombination current of a grain boundary network is therefore given by the
sum of its individual contributions, which can be weighted by a statistical
distribution of grain boundary parameter values. The description of the orientation-dependence and the network behavior of grain boundary recombination completes our model.  These advances expand the applicability of the model from idealized, artificial geometries to real materials.  Our model therefore provides a missing link between nanoscale characterization of the distribution of grain boundary properties and
their impact on a real device open-circuit voltage $V_{\rm oc}$.

\bigskip

\begin{figure}[t]
\centering
 \includegraphics[width=0.49\textwidth]{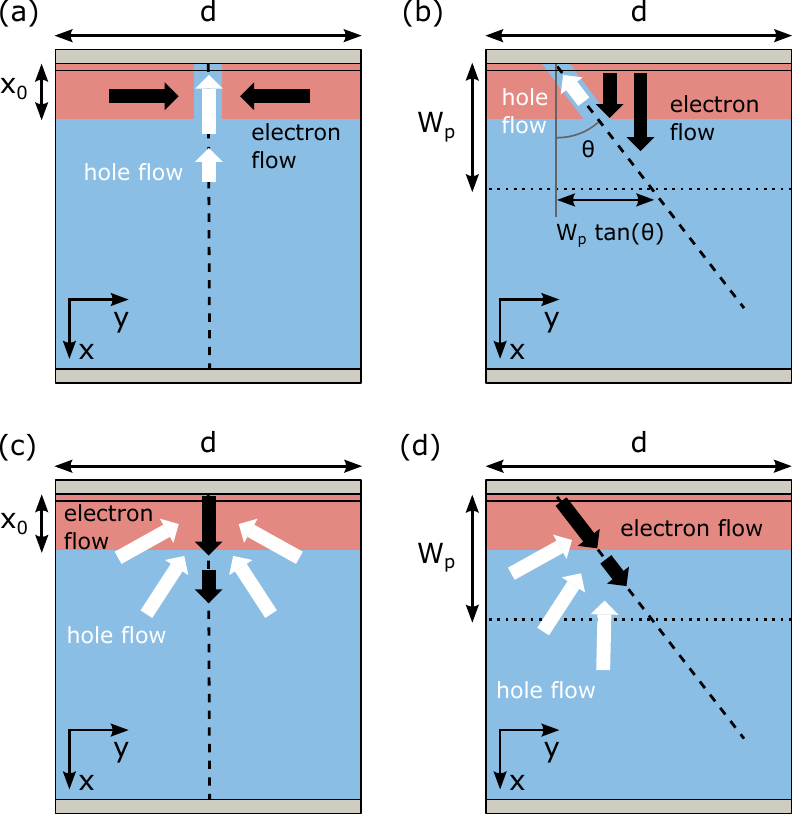}
 \caption{\label{colors} Schematics of the hole and electron particle
 currents. a), b) In a $p$-type grain boundary, for
 $\theta=0\degree$ and $\theta > 0$ respectively. c), d) In the
 high-recombination regime, for $\theta=0\degree$ and $\theta > 0$
 respectively. $W_p$ is the grain interior depletion region width, $x_0$ is the point where
 hole and electron concentrations are equal in the grain interior. Regions in
 blue and red are respectively $p$-type and $n$-type.}
\end{figure}

We begin with a description of the charge transport of a single grain boundary with one
end near the metallurgical $pn^+$ junction, and oriented an angle $\theta <
90\degree$, as shown in Figure~\ref{colors}b. We consider grain boundaries which do not make direct connections with the contacts.  The physical picture we describe here is based on Ref.~\citenum{Gaury2016GB},
which provides more details.  Informed by numerical simulations, we first posit that the grain boundary orientation primarily affects the length over which recombination occurs along the grain boundary core. That is, $\lambda$ of Eq. (\ref{form}) is assumed to be the only $\theta$-dependent factor.  We start with some definitions.  $W_p$ is defined as the $pn^+$ junction depletion width in the grain interior, {\it i.e.} in a region where the electrostatic potential is unperturbed by the grain boundary (see Figure~\ref{colors}d).  $x_0$ is defined as the position in the grain interior where $n=p$ (in equilibrium), such that $n>p$ for $x<x_0$ (see Figure~\ref{colors}c).  The primary quantity of interest is the grain boundary recombination, which occurs at defects located at the grain boundary core.  For $n$-type or $p$-type grain boundaries, the grain boundary defect recombination is set by the minority carrier concentration, while for high-recombination grain boundaries, both electron and hole density control the recombination. We summarize the behavior of the system for the three cases below.

For an $n$-type grain boundary, recombination is determined by holes.  Ref. \citenum{Gaury2016GB} shows that holes flow from the bulk of the $p$-type grain interior into the grain boundary core and recombine.
The majority of the grain boundary length $L_{\rm GB}$ is embedded in $p$-type bulk, so
holes are available for transport into the grain boundary from
the grain bulk over approximately the entire grain boundary length, independent of the orientation $\theta$.
The high defect density fixes the quasi-Fermi levels and
the electrostatic potential over the entire length of the grain boundary, as
shown in Figure~\ref{pot}.  Electron and hole densities are therefore uniform
along the grain boundary, leading to uniform recombination along the entire
grain boundary length $L\GB$. Hence we have $\lambda=L\GB$.

The recombination in a $p$-type grain boundary is
determined by electrons, which flow from the $n$-type grain interior into the grain
boundary core. For a perfectly columnar ($\theta=0\degree$) grain boundary,
the electrons flow into the grain boundary for $x<x_0$, as shown in Figure~\ref{colors}a.  The recombination
is therefore concentrated within the $n$-region of the $pn^+$ junction
depletion region, and is uniform for $x<x_0$.
As the grain boundary is tilted ($\theta\neq0\degree$), a larger section of the grain boundary is exposed to electrons coming
from the nearby $n$-contact, as shown in Figure~\ref{colors}b. This increased
exposure expands the region of uniform recombination, which leads to a longer
recombination region $\lambda$.  We find the appropriate form for the increase in recombination length due to grain boundary tilting is $\lambda = x_0 + W_p\tan(\theta)$, where $W_p \tan(\theta)$
represents the horizontal cross section of the segment of the grain boundary
exposed to the electron flow in the depletion region.

Additional recombination occurs as electrons diffuse along the grain boundary, increasing $\lambda$.
Note that electron transport is not confined to the grain boundary dislocation core (which is of atomic scale), but is spread out over the depletion width surrounding the grain boundary core. We denote the length scale for electron confinement near a grain boundary by $L_{\mathcal{E}}$; for default material parameters (given in Table S2 of the Supporting Information) and moderate grain boundary potentials ({\rm e.g} 250 eV), $L_{\mathcal{E}}$ is on the order of $0.2~\mu{\rm m}$. The effective lifetime of confined electrons is then given by $L_{\mathcal{E}}/S$, where $S$ is the effective grain boundary recombination velocity, and their diffusion length $L_n$ is $\sqrt{D_{\GB}^e L_{\mathcal{E}}/S}$, where $D_{\GB}^e$ is the diffusivity of the confined electrons (which may be reduced from the bulk value due to disorder at the grain boundary core). For large surface recombination velocities, diffusion lengths of confined electrons are small, and additional recombination away from the $pn$ junction depletion width is negligible.  For low surface recombination velocities, diffusion lengths of confined electrons are large and recombination is uniform along the entire grain boundary. The length of the recombination region in these two limits therefore reads
\begin{align}
    \label{lambda_ptype1}
    \lambda &= x_0 + W_p \tan(\theta) &{\rm for}~L_n \ll L\GB \\
    \lambda &= L\GB &{\rm for}~L_n \gg L\GB
    \label{lambda_ptype2}
\end{align}
Equation~(\ref{lambda_ptype1}) is valid as long as $\theta$ is such that
$\lambda<L\GB$; $\lambda=L\GB$ otherwise.  Eq. (10) of the Supporting Information gives the general expression for $p$-type grain boundary recombination for a general value of $L_n$.

The high-recombination regime of the perfectly columnar grain boundary occurs at
sufficiently high applied voltage so that both electron and hole densities are
of comparable magnitude. In this case, grain boundary recombination is the
result of both electron and hole currents flowing into the grain boundary core.
Both carrier types are available only in the vicinity of the depletion region,
so that currents flow as depicted in Figure~\ref{colors}c. Holes flow towards the
$pn^+$ junction depletion region to recombine with electrons flowing along the
grain boundary core.  The recombination is therefore peaked
at a ``hotspot'' in the depletion region.~\cite{Gaury2016GB}  As the grain boundary is tilted, a longer
section is exposed to hole flow in the depletion region, as shown in
Figure~\ref{colors}d. This larger exposure increases the
recombination region length $\lambda$ in a manner similar to the previous $p$-type grain boundary case: $\lambda=W_p/2 \tan(\theta)$.
Beyond this ``hotspot'' region, electrons diffuse in a one-dimensional
motion along the grain boundary core, as in the $p$-type grain boundary case described above.
We find that the recombination region in the high-recombination regime reads
\begin{align}
    \label{x0'1}
    \lambda &= \frac{W_p}{2} \tan(\theta) & {\rm for}~L_n' \ll L\GB \\
    \lambda &= L\GB & {\rm for}~L_n' \gg L\GB
    \label{x0'2}
\end{align}
where $L_n'$ is the diffusion length of grain boundary-confined electrons in this regime. Equation~(\ref{x0'1}) is
valid as long as $\lambda < L\GB$, $\lambda=L\GB$ beyond that point.  Equation (14) of the Supporting Information gives the formula for high-recombination grain boundary current for a general value of $L_n'$.

\begin{figure}[t]
\centering
 \includegraphics[width=0.49\textwidth]{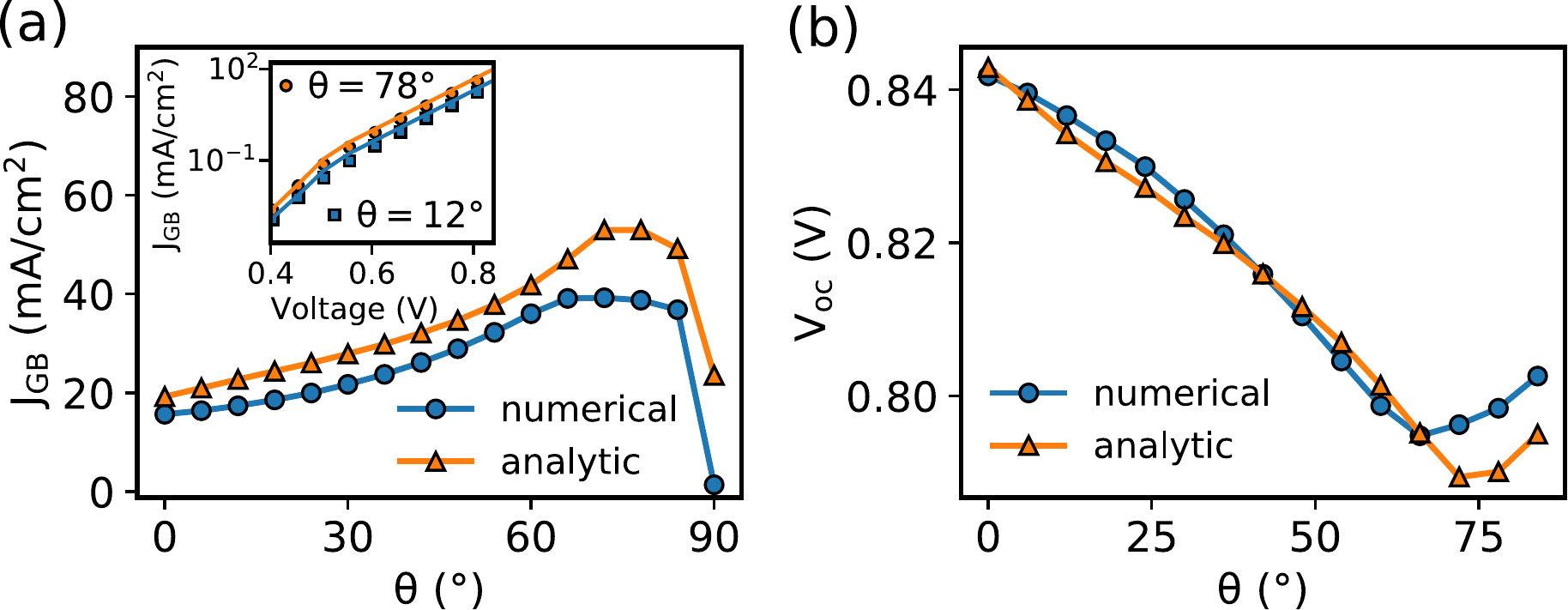}
 \caption{\label{angular} (a) Angular dependence of the current at fixed $V=0.8~{\rm V}$  Orange triangles denote analytic result, and blue dots denote numerical result.  The grain boundary contains a single donor and acceptor defect at energy $E\GB =0.53~{\rm eV}$ (measured from valence band edge), leading to downward band bending of $0.25~{\rm eV}$, with defect density $\rho\GB=10^{14}~{\rm cm^{-2}}$, recombination velocity $S\GB=10^5~{\rm cm/s}$, and length $L\GB=2.8~\mu{\rm m}$.  Other parameters are given by default values. Inset shows the current-voltage for two different values of $\theta$.  Symbols correspond to numerical data, full lines are analytic predictions.  (b) The analytical and numerically computed open-circuit voltage as a function of grain boundary orientation (same parameters as in (a)), for an incident photon flux of $2.5\times 10^{17}~{\rm cm^{-2}s^{-1}}$ and absorption length of $2.3\times 10^{4}~{\rm cm^{-1}}$.}
\end{figure}

To verify the accuracy of the above expressions, we compare our analytical prediction with the results of numerical simulation.  Details of the simulation software (along with its source code and a standalone executable) can be found in Ref. \citenum{gaury2018sesame}.
The simulation parameters are given in Table S2 of the Supporting Information and in the caption of Fig. \ref{angular}. Figure \ref{angular}(a) shows a comparison of the numerically computed grain boundary recombination current (blue dots) and the analytical predictions (orange triangles) as a function of grain boundary orientation for a fixed applied voltage ($0.8~{\rm V}$).  We find good agreement until the grain boundary becomes nearly completely horizontal, at which point the numerically computed current drops nearly to zero.  We find that our model does not describe this full blocking configuration, however it remains accurate at $\theta=85^\circ$.

We next consider the open-circuit voltage $V_{\rm oc}$.  Our model describes the dark forward bias current, so its applicability to $V_{\rm oc}$ relies on the superposition principle. At high forward bias, the carrier densities are large enough so that
quasi-Fermi levels and the electrostatic potential have negligible differences
with those in the dark.~\cite{Tarr1980} Because of the high recombination rate of grain boundaries, we find that
the current-voltage relation of the $pn^+$ junction under illumination
is given by the sum of the short circuit current $J_{\rm sc}$ and
the dark current only near $V=V_{\rm oc}$ (this superposition principle does not
apply in our system at lower voltages).  We use the analytical model to predict $V_{\rm oc}$ by shifting the analytical dark $J(V)$ curve by the numerically computed $J_{\rm sc}$.
Figure \ref{angular}(b) shows a comparison between the resulting $V_{\rm oc}$ for analytic and numerical models as a function of grain boundary orientation.  In both cases we find good agreement, demonstrating the accuracy of the analytical form for the open-circuit voltage.  However we find a large discrepancy for a horizontal grain boundary $\theta=90^\circ$, where the analytic $V_{\rm oc}$ is less than the numerical value by 0.09 V for the same reasons as the dark current discrepancy given above.  We omit this data point in Fig. \ref{angular}b.  The general form of the open-circuit voltage can be found by setting the general form for the high-recombination grain boundary current (given in its full explicit form in Eq. (14) of the Supporting Information) equal to $J_{\rm sc}$ and solving for $V$.


With a description of the isolated grain boundary recombination current as a
function of electrical and geometrical properties, we move on to consider
systems with multiple grain boundaries. It is not clear \textit{a priori} that the picture of
isolated grain boundary recombination is relevant to an arbitrary configuration
of grain boundaries. To address this question, we first reiterate the model conditions and assumptions: 1. Grain boundaries are positively charged.   2.
Grain boundaries have ``high'' defect density (see Supporting Information Eq. (5) for a precise criterion). 3. The hole quasi-Fermi level varies by less than $V_T$.  The validity of assumption 3 is the most difficult to assess.  This assumption may fail as a result of poor hole transport, due
to low hole mobility and/or low hole carrier concentration.  For networks of grain boundaries, the wide variety
of possible system geometries and parameters make it difficult to derive a precise and
general set of criteria for the validity of assumption 3 and the applicability of the analytical model. In lieu of such a
criterion, we numerically explore parameter space to explicitly find the domain of parameter values
for which the analytical model applies.

\begin{figure*}
\centering
 \includegraphics[width=1.0\textwidth]{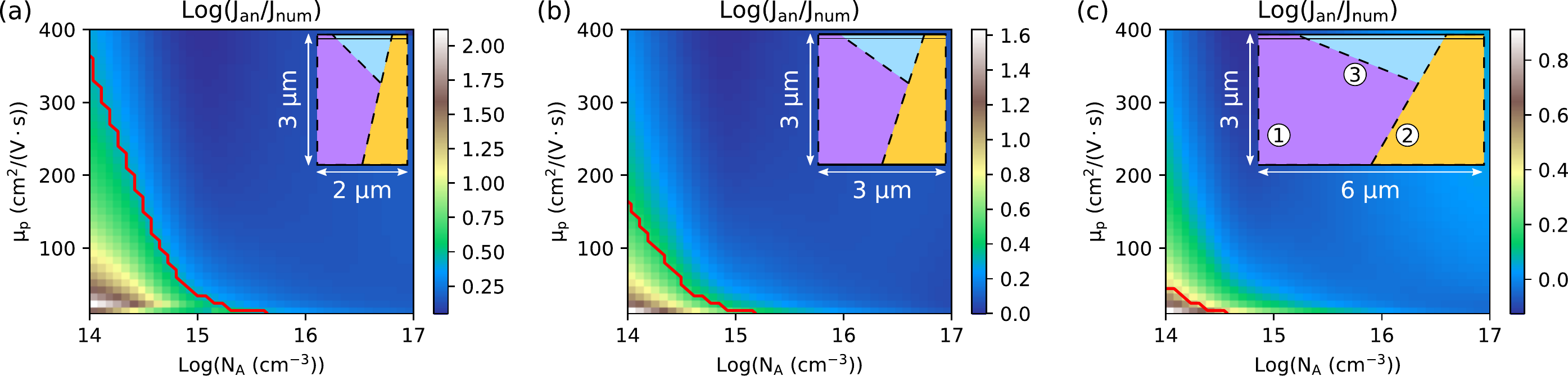}
 \caption{Contour plot of ratio of analytic to numerically computed forward bias dark current at $V=0.8~{\rm V}$.  Note that color scale is given on a ${\rm Log}_{10}$ scale. The red line delimits parameter space at which the ratio $J_{\rm an}/J_{\rm num} = \exp(1) \approx 2.7$.  The electron mobility is fixed to be $8\times$ the hole mobility.  Inset shows schematic of system geometry.  Grain boundaries labeled 1, 2, and 3 have built-in potential values of $(0.71,~0.25,~0.14)~{\rm eV}$ and recombination velocities of $\left(10^4,~5\times10^5,~10^5\right)~{\rm cm/s}$, respectively.  Top and bottom edges of schematics represent $n$ and $p$ contacts, respectively.  Left and right edges are modeled with periodic boundary conditions.  Note grain boundary 1 is located at the left edge.
  \label{validity}
}
\end{figure*}

We consider a system with 3 grain boundaries as depicted in the insets of Fig. \ref{validity}.  The defect
energy levels of grain boundary 1, 2, and 3 lead to downward band bending values of $(0.71,~0.25,~0.14)~{\rm eV}$, respectively (see inset of Fig. \ref{validity}(c) for grain boundary labels).
Fig. (\ref{validity}) provides a comparison of the analytical model with the numerical simulations
as a function of hole mobility $\mu_p$ and hole doping $N_A$ for three different grain sizes (fixed by system size $L_y$). We choose these three parameters because they most strongly determine the applicability of the analytical model.
We plot the ratio of the analytically predicted to numerically computed dark current at a fixed forward bias voltage $V=0.8~{\rm V}$.
The red lines delimit the region in parameter space for which the ratio is greater/less than $e\approx 2.7$. We
find that dark current ratio values of less than 2.7 correspond to systems for which the analytically predicted $V_{\rm oc}$ deviates
from the numerically simulation value by less than the thermal voltage $25~{\rm mV}$ (see Fig. S1 of the Supporting Information).
As expected, the factors which limit hole transport: low hole mobility and/or low hole carrier concentration due to low hole doping or depleted grains (i.e. grains smaller than the grain boundary depletion width) cause the analytical model to fail.  However we find that the analytical model accurately describes the numerical simulation for a wide range of system parameters.

The precise limits of parameter space for which our analytical model applies depends on the details of the grain boundary geometry and defect parameters.
For example, if we reduce the electrostatic band bending of grain boundary 1 from 0.71 eV to 0.21 eV, then we find
the region of analytic model applicability increases slightly (see Fig. S2 of the Supporting Information).  This can be expected: a decreased
grain boundary built-in potential decreases the depletion width surrounding the grain boundary, so that
hole carriers are less depleted and our assumption of facile hole transport is more easily satisfied.
However, we find that the boundaries presented in Fig. \ref{validity} give a fairly representative indication of the analytical model's domain
of validity.  We note that for a grain size of $\approx 1.5~\mu{\rm m}$, the analytical
model can be applied for parameters typical of CdTe absorbers: $\mu_p=40~{\rm cm^2/\left(V\cdot s\right)}$ and $N_A=4\times 10^{14}~{\rm cm^{-3}}$.

\begin{figure*}
\centering
 \includegraphics[width=0.8\textwidth]{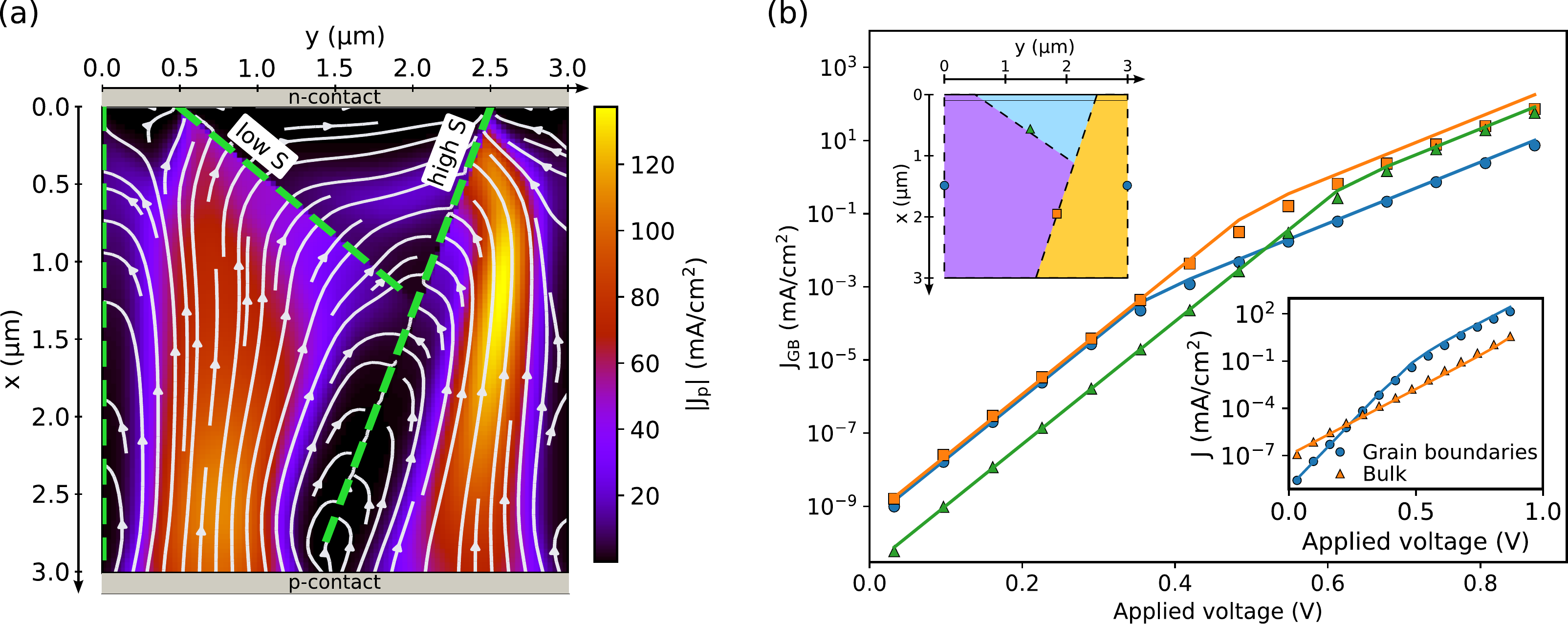}
 \caption{\label{manyGBs} Dark recombination current of a grain boundary
 network.
 a) Hole
 current map for dark current at $V=0.8~{\rm V}$.
 White lines correspond to the hole current field. The effective surface
 recombination velocity of the grain boundary on the left (``low S'') is $5\times $ less than the one of the grain boundary on the right (``high S'').
 b) Grain boundary dark recombination current as a function of voltage for a
 structure with multiple grains/grain boundaries as
 depicted in the upper inset. Symbols are numerical data, full lines are
 analytic predictions. Grain boundaries characteristics are same as in Fig. \ref{validity}. Lower
 inset: sum of the grain boundaries dark recombination currents (blue dots) and
 bulk recombination current (orange triangles). The analytic predictions for the
 bulk recombination current is given by  Equation~(28) and (34) of
 Ref.~\citenum{Gaury2016GB}. Simulation parameters
 are in Table~S2 of the Supporting Information.
}
\end{figure*}

We consider the behavior of a specific system in more detail in Fig. \ref{manyGBs}.  For this simulation we use default parameter values ($\mu_p=40~{\rm cm^2/\left(V\cdot s\right)}$, $N_A=4\times 10^{14}~{\rm cm^{-3}}$).  We first show the field
lines of the hole currents in Fig. \ref{manyGBs}a. When the hole
currents transverse to both sides of the grain boundary core are equal and opposite (see $x>2~\mu\rm m$), the
transverse hole current vanishes at the grain boundary core. For $x<2~\mu\rm m$,
only one side of the grain boundaries has direct access to the
$p$-contact. In this case, hole currents can partially go through grain boundaries,
as seen around the left grain boundary. A fraction of the incoming holes
recombine at the grain boundary core, so that hole currents on both sides are not equal.
Not surprisingly, holes that did not recombine are then attracted preferentially
to the grain boundary with the highest surface recombination velocity (``high
S'' on Figure~\ref{manyGBs}a).

\bigskip
\noindent

In Figure \ref{manyGBs}, we plot the numerically computed recombination current of the three grain boundaries separately (symbols),
together with the analytical predictions (solid lines).  In this case, the analytical theory overestimates the numerically computed current by approximately a factor of 2 at high applied voltage.  At low applied voltages, all grain boundaries are either $n$-type or $p$-type,
with ideality factor of 1. The transition to the high-recombination regime is revealed by the change of slope,
corresponding to an ideality factor of 2.  The lower inset of Figure~\ref{manyGBs}b compares the total grain boundary and bulk recombination
currents (the latter was computed in Ref.~\citenum{Gaury2016GB}). In addition
to its larger amplitude, the grain boundary's recombination current exhibits
change of slope. For most of the applied
voltages, the bulk recombination is proportional to $\exp(qV/(2k_BT))$ as given
by the $pn^+$ junction depletion region recombination.~\cite{Gaury2016GB} Because
of the variety of grain boundary properties in our geometry, the grain
boundary most dominating the dark current changes with applied voltage
leading to multiple changes of slope between $\exp(qV/k_BT)$ and
$\exp(qV/(2k_BT))$. This feature distinguishes grain boundary
recombination from $pn^+$ junction depletion region recombination.

\begin{figure*}
\centering
 \includegraphics[width=\textwidth]{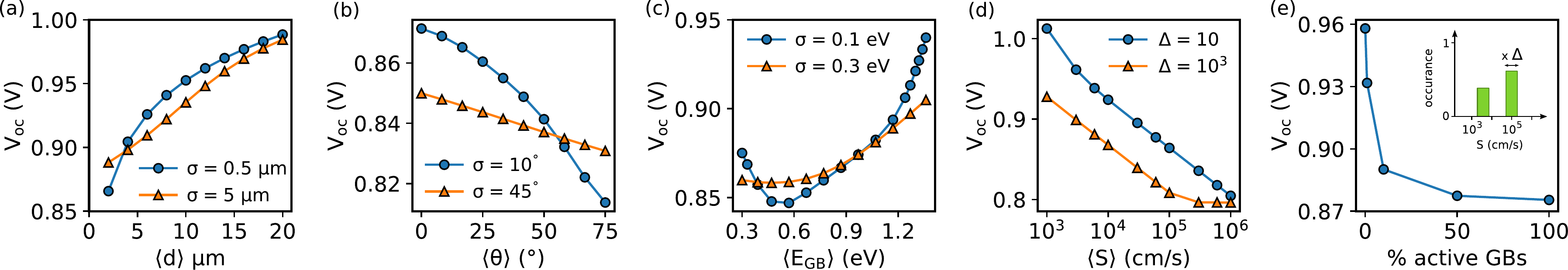}
 \caption{\label{voc} Open-circuit voltage for the system described in
 Fig.~\ref{geometry} under a photon flux $2.5\times 10^{17}~{\rm cm^{-2}s^{-1}}$. The
 absorption coefficient is $2.3\times 10^4~\rm cm^{-1}$. $\sigma$ is the standard
 deviation of a Gaussian distribution for a given parameter. $\sqrt{\Delta}$ is
 the geometric deviation of a uniform distribution.
 a) $V_{\rm oc}$ as a function of the average grain size $d$.
 b) $V_{\rm oc}$ as a function of the average grain boundary angle
 $\theta$ with respect to the normal to the $pn^+$ junction.
 c) $V_{\rm oc}$ as a function of the average neutral point of the gap
 state configuration $E\GB$.
 d) $V_{\rm oc}$ as a function of the (geometric) average grain
 boundary effective surface recombination velocity $S$.
 e) $V_{\rm oc}$ as a function of the percentage of active grain
 boundaries for the bimodal distribution shown in inset.
 Only one parameter is
 varied on each plot, the others are fixed to:
 $d=2.3~\mu {\rm m},~\theta=0\degree,~E\GB=1~{\rm eV},~S=10^5~\rm
 cm/s$.}
\end{figure*}

\bigskip
\noindent

Having established the conditions for which the analytical model describes the behavior
of multi-grain boundary systems, we proceed with an analysis of the impact of grain boundary
inhomogeneities on the open-circuit voltage using the analytical model alone. We consider an ensemble of ``samples'', each with its own distributions of grain size,
grain boundary orientation, gap state configuration and surface recombination velocity. For a probability distribution
$P$ of the random parameter $X$ with mean $\mu$ and standard deviation $\sigma$, the average grain
boundary dark current reads
\be
    \langle J\GB(V) \rangle = \int \mathrm{d}X\ J\GB(V, X)
    P(X; \mu, \sigma)
    \label{integral}
\ee
and the open-circuit voltage is found by solving $\langle J\GB(V_{\rm
oc})\rangle = J_{\rm sc}$.

Assuming equal short-circuit currents among all ``samples'', Figure~\ref{voc}
shows the open-circuit voltage as a function of the mean of the distribution of
a grain/grain boundary property.  Note that a discussion on the device
efficiency, which is beyond the scope of this work, should also include the
impact of grain boundary properties on $J_{\rm sc}$. The integral \eq{integral} was computed
numerically with the general form for $J\GB$ given by \eq{form}. We used
Gaussian distributions for Figure~\ref{voc}a,b,c, with small (lines with dots) and
large (lines with triangles) deviations from
the distribution mean. Because surface recombination velocities vary by orders
of magnitude, we chose a uniform distribution over the interval $[\langle S
\rangle/\sqrt{\Delta}, \sqrt{\Delta}\langle S \rangle]$ for Figure~\ref{voc}d,e,
where $\sqrt{\Delta}$ is the geometric deviation of the distribution.  Wide
(narrow) distributions represent strong (low)
inhomogeneities in the grain/grain boundary property.

$V_{\rm oc}$ varies
logarithmically with grain size as shown by the dots in Figure~\ref{voc}a,
so only large variations produce significant change in the
open-circuit voltage. The weak influence of grain size on open-circuit voltage
was observed with photoluminescence measurement on CdTe solar
cells.~\cite{Metzger2003}  Figure~\ref{voc}b shows that
grain boundaries forming low angles with the normal to the $pn^+$ junction are always more
favorable, almost regardless of the inhomogeneities. The reduction of
open-circuit voltage at large angles results from the increase of the length of
the recombination region as shown by \eq{x0'1}. The gain of approximately
$50~\rm mV$ in $V_{\rm oc}$ (around 5~\% increase in typical values of
$V_{\rm oc}$ in CdTe)
 encourages the engineering of a material growth
process that increases the proportion of quasi-perfectly columnar grains.
Efforts in this direction have been
reported.~\cite{Luschitz2009,Spalatu2015}

Figure~\ref{voc}c shows that gap state neutral points $E\GB$
close to a band edge generally give better $V_{\rm oc}$ than midgap values.
Note that the grain boundary built-in potential attracts photogenerated
electrons to the grain boundary core, resulting in enhanced recombination when
holes are majority carriers there. The short circuit current is therefore more
reduced by $p$-type than $n$-type grain boundaries (see Supporting Information), which
Figure~\ref{voc}c does not show. Thus, only high values of built-in potential
(i.e.,  $\langle E\GB \rangle > 1~\rm eV$) are favorable for the open-circuit
voltage.  Experimentally, the neutral point of the gap state distribution
determines the amplitude of the grain boundary built-in potential $V\GB$. At
thermal equilibrium and in the limit of high defect density of states, this
relation reads $qV\GB \approx E\GB - E_F$ ($E_F$ is the bulk Fermi energy).
Increasing the spread of $E\GB$ results in more gap state neutral levels around
midgap.  The probability to access midgap states being similar for both
electrons and holes, such states provide higher recombination currents than
states near the band edge, hence further reducing $V_{\rm oc}$. These results
are consistent with the observation that the standard CdCl$_2$ treatment of CdTe
increases the built-in potential around grain
boundaries,~\cite{tuteja2016direct} possibly leading to majority carrier type
inversion under suitable conditions.~\cite{li2014grain, Lukas2014}

The reduction of the open-circuit voltage by the recombination strength is
quantified in Figure~\ref{voc}d,e, where $S$ is the effective surface
recombination velocity.  We observe the expected logarithmic
dependence of $V_{\rm oc}$ on recombination velocity in Figure~\ref{voc}d. The
highest surface recombination velocities of the two distributions considered in
the figure differ by a factor 10. This is consistent with the
 difference in $V_{\rm oc}$ of about $50~\rm mV$ ($\approx k_BT/q\ln(10)$) between the two
distributions.
This shows that the largest value of surface
recombination velocity in the sample determines $V_{\rm oc}$.  Note that because
we only allow recombination velocities below the thermal velocity, $V_{\rm oc}$
saturates for $\langle S \rangle > 10^5~\rm cm/s$ in the case of $\Delta=10^3$.
The control of the open-circuit voltage by the most deleterious grain boundaries
is further illustrated in Figure~\ref{voc}e.
A cathodoluminescence study of CdTe grain boundaries revealed that approximately
$60~\%$ of the boundaries were active recombination centers.~\cite{Moseley2015}
Here we show that even a small proportion of active recombination
centers is sufficient to degrade $V_{\rm oc}$.
We assumed a bimodal distribution of
active and inactive grain boundaries with low ($S_1=5\times 10^3~\rm cm/s$) and high
($S_2=10^5~\rm cm/s$) average recombination velocities, as shown in inset.  In this
instance, we find that a proportion of only $10~\%$ of active grain boundaries
gives an open-circuit voltage equivalent to that of a system fully saturated with
active grain boundaries. This observation is best understood by assuming that
only surface recombination velocities $S_1$ and $S_2$ are present in the sample
(i.e., a probability distribution with two Dirac delta functions). In this case
the average surface recombination velocity across the sample is
\be
    \langle S \rangle = (1-p) S_1 + p S_2,
    \label{pS}
\ee
where $p$ is the proportion of active grain boundaries.
As $S_1$ and $S_2$ are separated by orders of magnitude, \eq{pS} shows that only
a small proportion of active grain boundaries (e.g., $p=0.1$) is sufficient to
dramatically shift the average recombination velocity towards high values.

Our last observation is the resilience of $V_{\rm oc}$ to moderate
inhomogeneities, a common feature among the wide parameter distributions.
In the high recombination regime (most relevant around $V_{\rm oc}$), the dark
recombination current varies only algebraically (not exponentially) with all the
grain boundary parameters. In turn, the open-circuit voltage varies
logarithmically with these parameters, leading to some tolerance towards
variations.

\bigskip
In this work, new analytical results were provided for the dark recombination
current of grain boundaries forming non-perfectly columnar grains. The model
accounts for the main features of grain boundaries: grain size, orientation, gap
state configuration and recombination strength. Generalizing
the isolated grain boundary picture to networks of grain boundaries is accomplished
by finding the parameter space for which contributions of grain boundary recombination
can be added independently. We find that for parameters relevant for many thin film photovoltaics,
this generalization is valid. Applying these results to random
distributions of grain boundary properties led to practical observations
regarding the open-circuit voltage. In particular, $V_{\rm oc}$ tolerates
moderately heterogeneous grain boundary properties.


To our knowledge, this work is the first fully analytical, quantitative
description of grain boundary networks. It contributes to bridge the gap between
experimental determination of grain boundary properties and their impact on the
device open-circuit voltage. The combination of this theory with nanoscale
measurements and first principle calculations can lead to a comprehensive
approach to improve the performance of thin-film solar cell technologies.


\bigskip
\noindent\textbf{Supporting information}

Explicit form for the analytical current-voltage relations; numerical simulations demonstrating that the analytic dark current-voltage relation accurately predicts the open-circuit voltage; additional numerical simulations demonstrating that the values of grain boundary built-in potential does not appreciably change the regime of validity for the analytical model; table of default simulation parameters

\bigskip
\noindent

\bigskip
\noindent\textbf{Acknowledgments}

\bigskip
\noindent
B.~G. acknowledges support under the Cooperative Research
Agreement between the University of Maryland and the National Institute of
Standards and Technology Center for Nanoscale Science and Technology, Award
70NANB14H209, through the University of Maryland.

\bigskip

\providecommand{\latin}[1]{#1}
\makeatletter
\providecommand{\doi}
  {\begingroup\let\do\@makeother\dospecials
  \catcode`\{=1 \catcode`\}=2 \doi@aux}
\providecommand{\doi@aux}[1]{\endgroup\texttt{#1}}
\makeatother
\providecommand*\mcitethebibliography{\thebibliography}
\csname @ifundefined\endcsname{endmcitethebibliography}
  {\let\endmcitethebibliography\endthebibliography}{}

\clearpage
\begin{center}
\noindent\textbf{Supporting information}
\end{center}

\bigskip
\noindent
\begin{table}[b]
\setlength{\tabcolsep}{0.1cm}
\begin{tabular}{cccc}
  \toprule
 Param. & $n$-type & $p$-type & high-recombination\\
  \midrule
  n  & $1$ & $1$ & $2$\\
  $E_a$  & $E\GB$ & $E_g-E\GB$ & $E_g/2$\\
   N& $N_V$ & $N_C$ & $\sqrt{N_CN_V}$ \\
   S & $S_p$ & $S_n$ &
   $\sqrt{S_nS_p}$\\
  $\lambda$  & $L\GB$ & \begin{tabular}{cc}
    $L\GB~\mathrm{for}~L_n \gg L\GB$ \\
    $x_0~\mathrm{for}~L_n \ll L\GB$ \end{tabular}&
    \begin{tabular}{cc}
    $L\GB~\mathrm{for}~L_n' \gg L\GB$ \\
    $L_n'~\mathrm{for}~L_n' \ll L\GB$ \end{tabular}
  \\
  \bottomrule
\end{tabular}
\caption{\label{general} Summary of analytical results for the grain boundary
recombination current for a continuum of donor and acceptor defect states. The
general form of the grain boundary dark current is $J\GB(V)=\lambda S/(2d) N
e^{-E_a/k_BT} e^{qV/(nk_BT)}$ where $S$ is an effective surface recombination velocity,
$\lambda$ is a length characteristic of the recombination region, $d$ is the
grain size, $N$ is an
effective density of states, $E_a$ is an activation energy, $n$ is the ideality
factor and $V$ is the applied voltage.  Each column corresponds to the regime in
which the grain boundary is depending on voltage. $L\GB$ is the length of the
grain boundary, $L_n$ and $L_n'$ are effective electron diffusion lengths.}
\end{table}
We use the model for charged grain boundaries of Ref.~\citenum{Gaury2016GB},
which is summarized here.  A single grain boundary is modeled as a two-dimensional plane
with a donor and acceptor gap states at equal energy $E\GB$.  This is a convenient model that
exhibits Fermi level pinning at a charge neutrality level \cite{taretto2008numerical}.  The corresponding grain boundary charge
density reads
\be
Q\GB = q \rho\GB \left(1-2 f\GB\right),
\label{QGB}
\ee
where $\rho\GB$ is a two-dimensional defect density. The defect state occupancy is \cite{PhysRev}:

\be
    f_{\rm GB} = \frac{S_n n\GB + S_p \bar
    p\GB}{S_n(n\GB+\bar n\GB) + S_p(p\GB+\bar p\GB)},
    \label{eq:fGBe}
\ee
where $n\GB$ ($p\GB$) is the grain boundary  electron (hole) carrier density, $S_n$
($S_p$) is the electron (hole) surface recombination velocity, $\bar
n\GB$ and $\bar p\GB$ are given by:
\begin{align}
\label{nbar}
    \bar n\GB &= N_C e^{\left(-E_g + E\GB\right)/k_BT}\\
    \bar p\GB &= N_V e^{-E\GB/k_BT},
\label{pbar}
\end{align}
where $E\GB$ is a defect energy level calculated from the valence band edge, $N_C$
($N_V$) is the conduction (valence) band effective density of states, $E_g$ is
the material bandgap, $k_B$ is the Boltzmann constant and $T$ is the
temperature.  The parameters $S_{n,p}$ and $\rho\GB$ are related to the
electron and hole capture cross sections $\sigma_{n,p}$ by $S_{n,p} =
\sigma_{n,p} v_t \rho\GB$, where $v_t$ is the
thermal velocity. In the present work we varied $S_{n,p}$ with fixed $\rho\GB$; this
corresponds to varying $\sigma_{n,p}$ accordingly.

We consider large defect densities of states such that the Fermi level is pinned
near the defect energy level $E\GB$ (which is the charge neutrality level for this model). This regime was
found to be reached for densities above the critical value

\be
    \rho_{\rm GB}^{\rm crit} = \frac{2}{q} \left(\frac{e+1}{e-1}\right) \sqrt{8q\epsilon N_A\left(E_{\rm GB}-E_F\right)}
    \label{critt}
\ee
where $E_F$ is the equilibrium Fermi energy and
$N_A$ is the doping density.  For default material parameters and grain boundary band bending of $0.5~{\rm eV}$,
$\rho_{\rm GB}^{\rm crit}$ is typically on the order of $5\times 10^{11}~\rm cm^{-2}$.

The diffusion length for electrons confined near the grain boundary depends on grain boundary type. We denote this with $L_n$ and $L_n'$ for $n$-type and high-recombination grain boundary, respectively. The relevant expressions are given below:
\begin{eqnarray}
L_n &=& 2\sqrt{D_n L_{\mathcal{E}}/S_n} \label{eq:Ln},\\
L_n' &=& \sqrt{8D_n L_{\mathcal{E}}'/\sqrt{S_n S_p}} \label{eq:Lnp},
\end{eqnarray}
where
\begin{eqnarray}
L_{\mathcal{E}} &=& V_T\sqrt{\frac{2\epsilon}{qN_A V_{\rm GB}^0}} \label{eq:LE},\\
L_{\mathcal{E}}' &=& \sqrt{\frac{2\epsilon V_T}{qN_A}} \label{eq:LEp},
\end{eqnarray}
where $V_T=k_BT/q$. $L_{\mathcal{E}}'$ is the length scale associated with the transverse electric field $\mathcal{E}_\perp$ of the grain boundary in the high  recombination regime: $L_{\mathcal{E}}=V_T/\mathcal{E}_\perp$.  In Eq. \ref{eq:LE}, $V_{\rm GB}^0$ is the equilibrium potential difference
 between grain boundary and grain interior.

We next provide the general expressions for the recombination current of $p$-type and high recombination grain boundaries (the expressions in the main text only show limiting values of $L_n/L\GB$ and $L_n'/L\GB$.  For $p$-type, the recombination is given by:
\begin{eqnarray}
J\GB(V) = \frac{S\GB n\GB}{2}  \exp\left(\frac{qV}{k_BT}\right) \left[ x_0 + W_p \tan\theta + L_n \left( 1 - \exp\left(-\frac{L\GB - x_0 - W_p\tan\theta}{L_n}\right)\right)\right]\label{eq:jGB_p}
\end{eqnarray}
As described in the main text, $x_0$ is the position where $n=p$ in equilibrium.  Its expression is:
\be
x_0 = W_p \left( 1 - \sqrt{1 - \frac{V_T}{V_{\rm bi}}\ln\left(\frac{N_D}{n_i}\right)}\right)
\ee
where $W_p$ is the bulk depletion width, and $V_{\rm bi}$ is the potential difference of the bulk $p$-$n$ junction:
\begin{eqnarray}
V_{\rm bi} &=& V_T \ln \left(\frac{N_A N_D}{n_i^2}\right) \\
W_p &=& \sqrt{\frac{2\epsilon V_{\rm bi}} {N_A}}
\end{eqnarray}
For high-recombination, the recombination is given by:
\begin{eqnarray}
J\GB(V) = \frac{S\GB n_i}{2}  \exp\left(\frac{qV}{2k_BT}\right) \left[\frac{ W_p \tan\theta}{2} + L_n' \left( 1 - \exp\left(-\frac{L\GB - (W_p\tan\theta)/2}{L_n'}\right)\right)\right] \label{eq:jGB_hr}
\end{eqnarray}
We have also checked that grain boundary networks with more complex defect electronic structure, such as a continuum of donors and acceptors (as described in Ref. \citenum{Gaury2017}), are accurately described by the approach we present here.

\begin{figure*}
\centering
 \includegraphics[width=.5\textwidth]{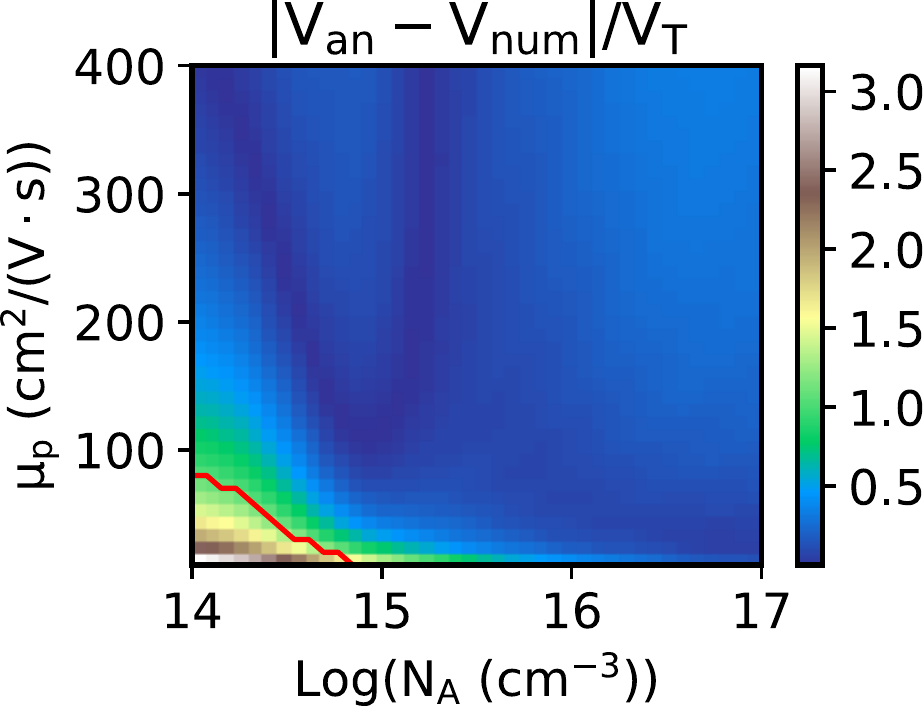}
 \caption{The difference in analytically predicted and numerically computed $V_{\rm oc}$ (scaled by thermal voltage $V_T\approx25~ {\rm mV}$) for the system given in Fig. 5b of the main text.  Red line indicates the parameter values for which this ratio is 1.  The electron mobility is fixed to be $8\times$ the hole mobility.  Note that the region of the applicability of the analytic model for $V_{\rm oc}$ is similar to the region of applicability for dark $J(V)$ (seen in Fig. 5b). \label{validity_voc}
}
\end{figure*}

Figure \ref{validity_voc} shows that the reliability of the analytical model's prediction for $V_{\rm oc}$ tracks its reliability for dark $J(V)$.  The red line indicating the region for which the analytical model predicts the numerically computed $V_{\rm oc}$ is similar to the region for which the analytical model predicts the numerically computed dark current to within a factor of $e\approx 2.7$ (see Fig. 5b).

\begin{figure*}
\centering
 \includegraphics[width=.7\textwidth]{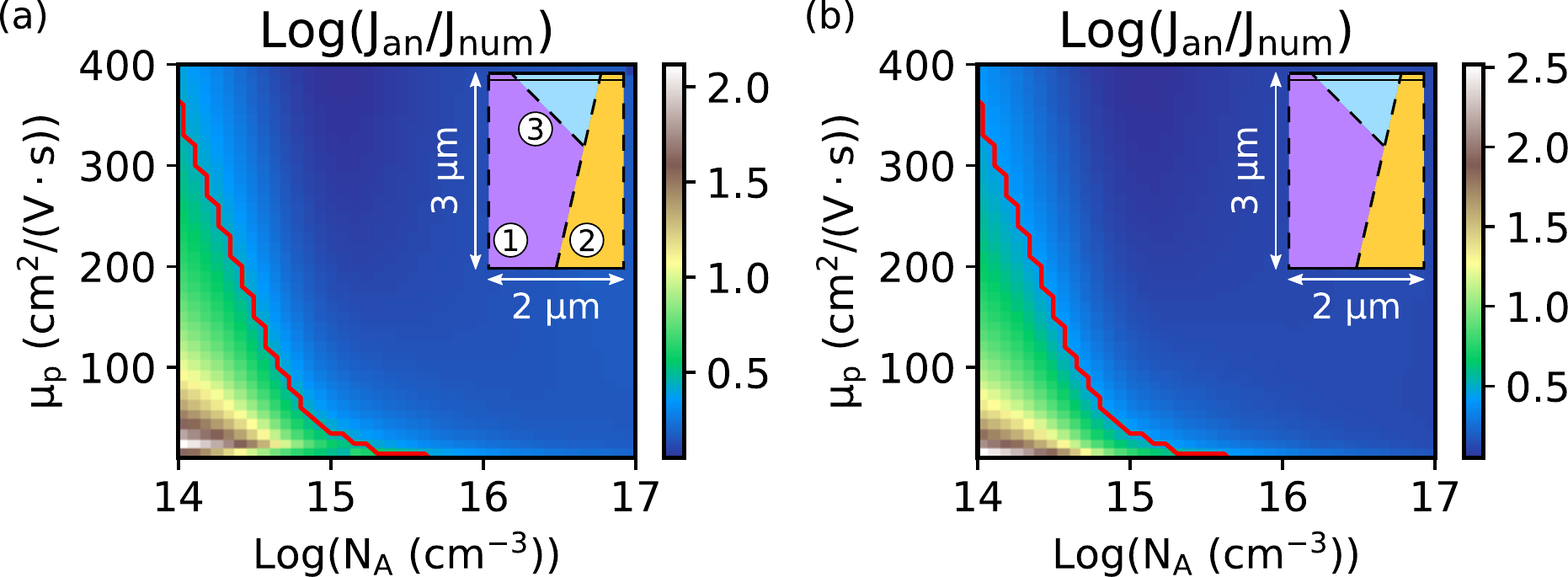}
 \caption{
 Contour plot of ratio of analytic to numerically computed forward bias dark current at $V=0.8~{\rm V}$.  Note that color scale is given on a ${\rm Log}_{10}$ scale. The red line delimits parameters space at which the ratio is $\exp(1)\approx 2.7$ (on a linear scale).  Inset shows schematic of system geometry.  For (a), grain boundaries labeled 1, 2, and 3 have built-in potential values of $(0.71,~0.25,~0.14)~{\rm V}$, while for (b) the built-in potential for grain boundary 1 is reduced to 0.21 ${\rm eV}$.  Recombination velocities for grain boundaries 1, 2, and 3 are $\left(10^4,~5\times10^5,~10^5\right)~{\rm cm/s}$, respectively for both (a) and (b).  Top and bottom edges of schematics represent $n$ and $p$ contacts, respectively.  Left and right edges are modeled with periodic boundary conditions.
 \label{smallvgb}
}
\end{figure*}

Figure \ref{smallvgb} shows the dependence of the analytical model performance on the grain boundary built-in potential.  We find that this parameter does not strongly influence the model reliability.  In general, the smaller the grain boundary built-in potential, the larger the regime of applicability, although the difference is quite small.

\bigskip
We tested our analytical predictions on numerical solutions of the
two-dimensional drift-diffusion-Poisson equations, solved using Sesame\cite{gaury2018sesame}. We used selective contacts,
so the hole (electron) current vanishes at $x=0$ ($x=3~\mu \rm m$). Periodic
boundary conditions were applied in the $y$-direction. Table~\ref{params} gives
a list of the material parameters used in these calculations.

\begin{table}[b]
\setlength{\tabcolsep}{0.1cm}
\begin{tabular}{llll}
  \toprule
  Param. & Value & Param. & Value \\ \midrule
  $L$ & $3~ \mu{\rm m}$ & $\epsilon$ & $9.4~\epsilon_0$ \\
  $d$ & $3~ \mu{\rm m}$  & $\tau_{n,p}$ & $10~{\rm ns}$ \\
  $N_C$ & $8\times10^{17}~{\rm cm^{-3}}$ & $S_{n,p}$ & $(10^4~\mathrm{to}~10^6)~{\rm cm/s}$ \\
  $N_V$ & $1.8\times10^{19}~{\rm cm^{-3}}$ & $N\GB$ & $10^{14}~{\rm cm^{-2}}$ \\
  $E_g$ & $1.5~{\rm eV}$  & $(\mu_{n},~\mu_p)$ & $(320,~40)~{\rm cm^2/\left(V\cdot s\right)}$ \\
  $N_D$ & $10^{17}~{\rm cm^{-3}}$ & $N_A$ & $4\times 10^{14}~{\rm cm^{-3}}$ \\
  \bottomrule\\
\end{tabular}
\caption{List of default parameters (Param.) for numerical simulations.\label{params}}
\end{table}

\providecommand{\latin}[1]{#1}
\makeatletter
\providecommand{\doi}
  {\begingroup\let\do\@makeother\dospecials
  \catcode`\{=1 \catcode`\}=2 \doi@aux}
\providecommand{\doi@aux}[1]{\endgroup\texttt{#1}}
\makeatother
\providecommand*\mcitethebibliography{\thebibliography}
\csname @ifundefined\endcsname{endmcitethebibliography}
  {\let\endmcitethebibliography\endthebibliography}{}

\end{document}